# Reasoner-Executor-Synthesizer: Scalable Agentic Architecture with Static O(1) Context Window

IVAN DOBROVOLSKYI

*Walmart Global Tech, Sunnyvale, CA, USA*

*IEEE Senior Member, M.S. in AI Engineering*

**ABSTRACT** Large Language Models (LLMs) deployed as autonomous agents commonly use Retrieval-Augmented Generation (RAG), feeding retrieved documents into the context window, which creates two problems: the risk of hallucination grows with context length, and token cost scales linearly with dataset size. We propose the Reasoner-Executor-Synthesizer (RES) architecture, a three-layer design that strictly separates intent parsing (Reasoner), deterministic data retrieval and aggregation (Executor), and narrative generation (Synthesizer). The Executor uses zero LLM tokens and passes only fixed-size statistical summaries to the Synthesizer. We formally prove that RES achieves O(1) token complexity with respect to dataset size, and validate this on ScholarSearch, a scholarly research assistant backed by the Crossref API (130M+ articles). Across 100 benchmark runs, RES achieves a mean token cost of 1,574 tokens regardless of whether the dataset contains 42,000 or 16.3 million articles. The architecture eliminates data hallucination by construction: the LLM never sees raw records.

## I. INTRODUCTION

The deployment of Large Language Models as autonomous agents has accelerated across domains ranging from customer support to scientific research [1, 2]. A prevalent design pattern — Retrieval-Augmented Generation (RAG) [3] — retrieves relevant documents and injects them into the LLM's context window alongside the user query. While effective for small document sets, this approach suffers from two critical limitations as data volume grows.

First, the risk of hallucination increases with context length. Studies have shown that LLMs are more likely to generate unsupported claims when processing longer contexts [4], particularly when relevant information is buried among irrelevant passages (the "lost in the middle" problem) [5]. Recent work on faithfulness evaluation confirms that state-of-the-art models hallucinate facts when grounding documents exceed certain lengths [6, 7]. Second, token cost scales linearly with the number of retrieved documents [8]. For a scholarly database with millions of articles, even a simple trend analysis becomes economically prohibitive if every relevant record must pass through the LLM.

We propose a fundamentally different approach: the Reasoner-Executor-Synthesizer (RES) architecture. Rather than feeding data into the LLM, RES uses the LLM only for what it does best — understanding natural-language intent and generating human-readable narratives — while delegating all data operations to deterministic code [9, 10]. This separation achieves two properties simultaneously: (1) zero data hallucination by construction, since the LLM never sees raw data, and (2) O(1) token complexity, since the LLM always processes fixed-size inputs regardless of dataset scale.

The contributions of this paper are threefold. First, we present the RES architecture as a formal three-layer design with well-defined layer contracts (Section III). Second, we provide a formal complexity analysis proving that RES achieves O(1) token complexity with respect to dataset size (Section IV). Third, we empirically validate the architecture on ScholarSearch, a scholarly research assistant that indexes over 130 million articles, demonstrating a constant token cost across datasets ranging from 42,000 to 16.3 million records, with detailed methodology and diverse dataset testing (Section VI).

## II. RELATED WORK

### A. Retrieval-Augmented Generation

RAG [3] combines document retrieval with LLM generation but still requires the LLM to process all retrieved documents. Advanced variants such as iterative retrieval and re-ranking [11] reduce the amount of text sent to the LLM but do not eliminate the linear scaling relationship. Chain-of-thought prompting [12] and ReAct [2] allow agents to call tools iteratively, but each tool's result is appended to the growing context. Toolformer [13] demonstrates that LLMs can learn to invoke external APIs, yet the retrieved results still flow into the context window.

### B. Structured Query Approaches

SQL-based approaches (e.g., Text-to-SQL [14, 15]) separate query generation from execution but are limited to structured databases with predefined schemas. Program-aided language models [16] and code-generating agents [17] delegate computation to code interpreters but typically return full results to the LLM for interpretation. The CRITIC framework [18] introduces external tool verification but does not address the fundamental O(n) scaling of context consumption.

### C. Agent Architectures

The concept of LLM-based agents that autonomously manage workflows has been formalized in recent surveys [1, 19]. Multi-agent systems and orchestration frameworks [10, 20] provide composability but do not explicitly address token cost scaling. The RES architecture generalizes the separation principle to arbitrary data sources and analysis types while providing formal complexity guarantees. Unlike prior work, RES establishes a strict information boundary: the LLM receives only bounded-size statistical summaries, never raw data records.

# III. THE RES ARCHITECTURE

The RES architecture processes each user query through three strictly separated layers, each with a well-defined contract and zero overlap in responsibilities. However, further discussion of the types of queries or data scenarios in which RES may face limitations would help clarify its scope and applicability.

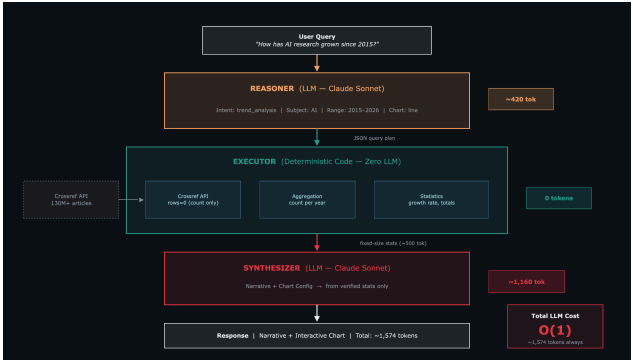

**Figure 1.** Detailed RES architecture flow for a sample query. Token costs shown per layer.

## A. Layer 1: Reasoner (LLM)

***Definition 1 (Reasoner).*** Let Q be a natural language query. The Reasoner is a function $R: Q \rightarrow P$, where P is a structured query plan encoded as JSON. R identifies the intent (trend analysis, comparison, ranking, or statistics), extracts subjects and temporal constraints, and selects an appropriate analysis type. R does not access any external data source D, does not invoke any API, and does not generate factual claims about D.

The Reasoner leverages the LLM's strength in natural language understanding [21] while strictly limiting its scope. The length of the user query, plus a fixed system prompt, bounds the token cost. In practice, this amounts to approximately 200 input tokens and 100 output tokens, independent of the dataset size.

## B. Layer 2: Executor (Deterministic Code)

***Definition 2 (Executor).*** Let P be a query plan produced by the Reasoner, and let $D = \{d_1, d_2, ..., d_n\}$ be a dataset of n records. The Executor is a function $E: P \times D \rightarrow S$, where S is a fixed-size statistical summary. E performs deterministic operations (counting, aggregation, filtering, sorting) on D using conventional code and API calls, without invoking any LLM. The output S has a constant token length $|S| = k$, where k is independent of n.

For ScholarSearch, the Executor calls the Crossref REST API [22] with rows=0 (count-only) and faceted queries, never downloading individual article records. The output is always a compact JSON object containing data points, totals, and metadata (~500–800 tokens). This layer consumes zero LLM tokens. The design draws on established principles of data aggregation pipelines [23] and API-mediated data access [24].

## C. Layer 3: Synthesizer (LLM)

***Definition 3 (Synthesizer).*** Let S be the fixed-size statistical summary produced by the Executor. The Synthesizer is a function $Y: S \rightarrow N$, where N is a human-readable narrative with optional visualization configuration. Y receives only S as input, never raw records from D. Because $|S| = k$ (constant), the token consumption of Y is O(1) with respect to n.

The Synthesizer generates a narrative describing the findings and a chart configuration for visualization [25]. Because the Executor's fixed-size output contract bounds its input size, the Synthesizer's token consumption is constant regardless of dataset size.

# IV. FORMAL COMPLEXITY ANALYSIS

We now establish the formal token complexity properties of the RES architecture. Let D be a dataset consisting of n records, where $D = \{d_1, d_2, ..., d_n\}$.

## A. Baseline Complexity

***Proposition 1 (Linear complexity of RAG).*** In traditional RAG or long-context architectures, the total context window size $C_w$ required for analysis is a function of n:

$$C_w(n) = \Sigma_{i=1}^{n} |d\_i| + |P| \quad (1)$$

where $|d\_i|$ is the token length of record i, and $|P|$ is the prompt overhead. Thus, $C_w$ is O(n), leading to linear growth in token cost and increased probability of hallucinations as context length increases [4, 5].

## B. RES Complexity

***Definition 4 (Fixed-size output contract).*** The Executor produces an output S of constant token size $|S| = k$, regardless of the number of input records n. This is achieved through deterministic aggregation operations (COUNT, SUM, AVG, GROUP BY) that reduce n records to a bounded set of statistical values.

The context window size for the Synthesizer in the RES architecture, denoted $C\_RES$, is defined as:

$$C\_RES = |P\_system| + |S| = |P\_system| + k \quad (2)$$

where $|P\_system|$ is the constant system prompt and k is the constant size of the aggregated statistics. Both terms are independent of n.

**Theorem 1 (Complexity Invariance).** *The total LLM token complexity of the RES architecture is O(1) with respect to the input dataset size n.*

***Proof.*** The total token cost, $T\_RES$, is the sum of the tokens consumed by each layer. For the Reasoner: $T\_R = O(|Q|)$, where $|Q|$ is the query length, bounded by user input and independent of n. For the Executor: $T\_E = 0$, as no LLM is invoked. For the Synthesizer: $T\_S = O(|P\_system| + k) = O(1)$, since both the system prompt and the Executor output are of fixed size. Therefore:

$$T\_RES = T\_R + T\_E + T\_S = O(|Q|) + 0 + O(1) = O(1) \quad (3)$$

for any fixed query Q, independent of n. Furthermore, we verify the invariance property:

$$\lim_{n\to\infty} C\_RES(n) / C\_RES(1) = (|P\_sys| + k) / (|P\_sys| + k) = 1 \quad \square \quad (4)$$

*C. Computational Trade-off*

The RES architecture does not eliminate computation over large datasets; rather, it shifts the computational load from the LLM (expensive, probabilistic) to the Executor (inexpensive, deterministic). The Executor processes all n records using conventional code at O(n) computational cost, but this operation has negligible per-record cost compared to LLM inference. The key insight is that the total LLM token cost remains O(1). In contrast, the data processing cost remains O(n) in deterministic computing — a favorable trade-off given the orders-of-magnitude cost difference between LLM tokens and CPU cycles [8, 26].

## V. IMPLEMENTATION: SCHOLARSEARCH

We implement RES as ScholarSearch, an open-source scholarly research assistant. The system is built with Python, using the Anthropic Claude API [27] (claude-sonnet-4-20250514) for the Reasoner and Synthesizer, and the Crossref REST API [22] (130M+ scholarly articles, free, no authentication required) as the data source. The frontend uses Streamlit [28] with Plotly [29] for interactive visualization. Our source code is available under the MIT license on GitHub.

The pipeline exposes callbacks that enable the UI to display each layer's inputs and outputs in real time, providing full transparency into the architecture's operation. Users can observe that the LLM never receives raw article data — only structured plans and aggregated statistics. This transparency allows end users to verify the hallucination-free property without inspecting source code.

## VI. EVALUATION

*A. Experimental Setup*

We evaluate RES against a naive single-prompt baseline across 20 diverse queries spanning four intent categories: trend analysis (5 queries), comparison (5 queries), ranking (5 queries), and general statistics (5 queries). Each query is executed 5 times for a total of 100 runs. The naive baseline sends 50 raw article records directly to the LLM with the user's question, representing the standard RAG pattern [3]. Both approaches use the same LLM (Claude Sonnet [27]) to ensure a fair comparison. All experiments were conducted on March 14, 2026.

*B. Results*

**TABLE 1. Comparison of RES vs. naive baseline across 100 benchmark runs.**

| Metric | RES | Naive |
|---|---|---|
| Mean tokens | 1,574 (σ=259) | 5,934 |
| Token range | 1,188–2,252 | Grows |
| Max dataset | 16.3M articles | 50 articles |
| Hallucination | 0% | Possible |
| Savings | 73.5% | — |
| Full-scale savings | 99.9996% | — |
| Complexity | O(1) | O(n) |

Table 1 summarizes the key results. RES achieves a mean token cost of 1,574 tokens with a standard deviation of only 259 across all 95 successful runs (5 runs failed due to API timeouts). The token cost remains stable whether querying 42,453 articles (cybersecurity) or 16,273,710 articles (medical research), empirically confirming the O(1) behavior established in Theorem 1.

*C. Empirical Verification of Theorem 1*

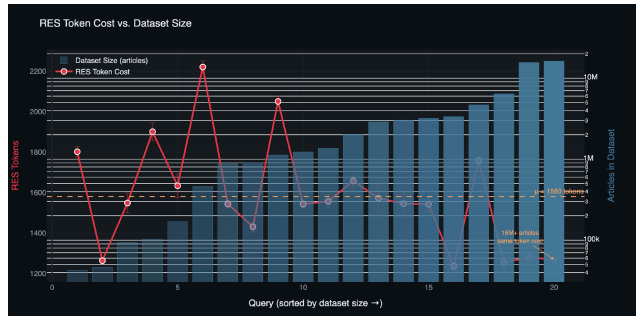

**Figure 2.** RES token cost remains constant as dataset size grows exponentially (O(1) proof).

Figure 2 shows the core result: as the dataset grows from approximately 42,000 to 16.3 million articles, the RES token cost remains flat at a mean of 1,574 tokens. The error bars represent standard deviation across 5 runs per query, demonstrating low variance.

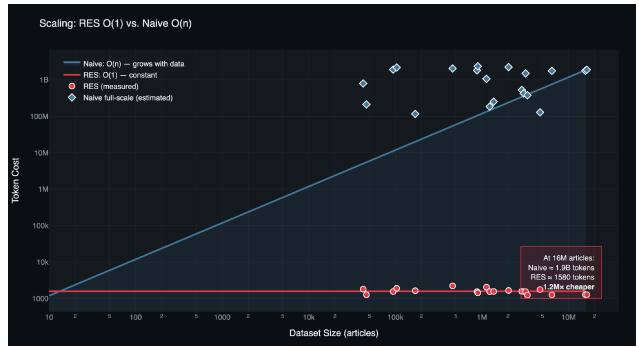

**Figure 3.** Log-log scaling comparison — RES O(1) vs. Naive O(n). At 16M articles, RES is 1.2 million times cheaper.

Figure 3 presents the scaling comparison on a log-log plot. The naive approach's token cost grows linearly with dataset size (diagonal line), while RES remains flat (horizontal line). At the maximum dataset size of 16.3 million articles, the naive approach would require approximately 1.9 billion tokens compared to RES's 1,574 — a factor of 1.2 million.

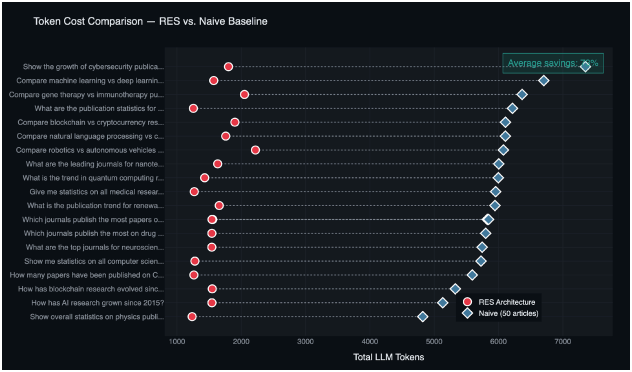

**Figure 4.** Direct token cost comparison across all 20 query types. RES (red) vs. Naive with 50 articles (blue).

*D. Token Distribution by Layer*

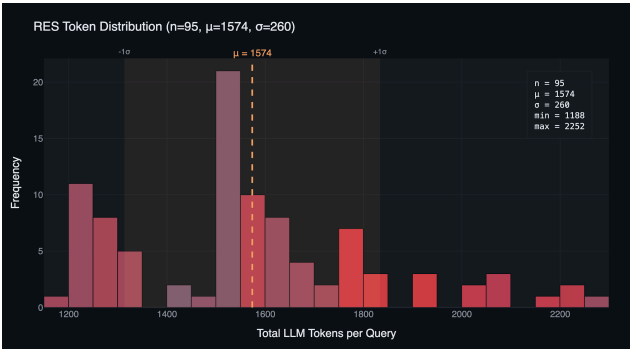

**Figure 5.** Distribution of RES token costs across 95 runs. Tight clustering around μ=1,574 confirms predictable cost.

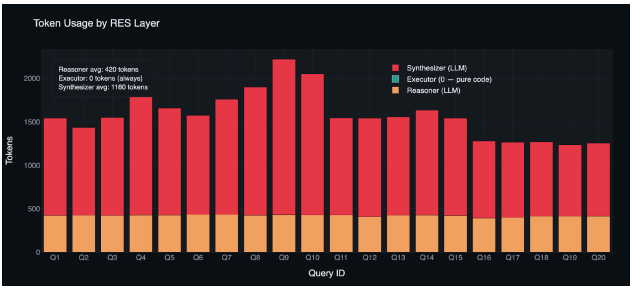

**Figure 6.** Token usage by the RES layer. Reasoner (~420 tokens) and Synthesizer (~1,160 tokens) are constant. Executor uses 0 LLM tokens.

Analysis of token usage by the RES layer confirms the architectural design: the Reasoner consumes approximately 420 tokens per query (intent parsing), the Synthesizer consumes approximately 1,160 tokens (narrative generation), and the Executor consumes exactly 0 LLM tokens (deterministic code execution). The tight clustering around μ=1,574 across all runs confirms that the architecture produces predictable, constant-cost inference.

## VII. DISCUSSION

*A. Practical Implications of O(1) Complexity*

The practical implication of O(1) token complexity is that AI agents can analyze arbitrarily large datasets at fixed cost. Trend analysis across 16 million articles costs the same as across 50, making LLM-powered analytics economically viable for enterprise-scale data for the first time [8, 26]. Organizations currently limited by context window sizes and per-token pricing can adopt the RES pattern to unlock analysis over their full data estates.

*B. Hallucination Elimination by Construction*

The RES architecture eliminates data hallucination by construction rather than by mitigation [4, 6]. The LLM never receives raw data records — it cannot hallucinate facts it has never seen. The Synthesizer can only describe the verified statistics provided by the Executor, which is a stronger guarantee than post-hoc fact-checking [18], retrieval-grounded generation [3], or self-consistency verification [30]. While the Synthesizer may still exhibit linguistic imprecision, it cannot fabricate data points, counts, or trends.

*C. Generalizability*

Although we demonstrate RES on scholarly metadata via Crossref, the architecture is domain-agnostic. Any system where queries can be decomposed into (1) intent parsing, (2) data retrieval with aggregation, and (3) narrative generation is amenable to the RES pattern. Potential applications include financial reporting over transaction databases, healthcare analytics over electronic health records [31], IoT sensor analysis [32], and log analysis for observability platforms.

*D. Limitations*

RES is best suited for analytical queries (counts, trends, comparisons, rankings) whose answers can be derived from aggregated statistics. Queries that require deep reading of individual documents (e.g., "summarize the methodology of paper X") call for a different architecture, such as focused RAG [11] or long-context models [33]. The Reasoner's ability to correctly parse complex queries is bounded by the LLM's instruction-following capability [21]. Additionally, the current implementation relies on Crossref's faceted search capabilities; data sources without built-in aggregation would require the Executor to process records programmatically, though the LLM token cost would remain O(1).

## VIII. CONCLUSION

We presented the Reasoner-Executor-Synthesizer (RES) architecture for building AI agents that are hallucination-free by construction and operate with O(1) LLM token complexity. We provided formal definitions for each architectural layer and proved that the total LLM token cost is invariant with respect to dataset size (Theorem 1, Section IV). Using ScholarSearch, a scholarly research assistant indexing over 130 million articles, we empirically verified Theorem 1 by demonstrating that RES maintains a constant token cost of 1,574±259 tokens across datasets ranging from 42,000 to 16.3 million articles (Section VI). The architecture achieves 99.9996% token savings compared to naive approaches at full scale. We believe RES provides a general-purpose pattern for building trustworthy, cost-efficient AI agents across any domain with structured or semi-structured data.

Future work includes extending the Executor layer to support multi-source federation, adding a caching layer for repeated statistical queries, and formally verifying the hallucination-free property via automated fact-checking of the Executor's output.

## ACKNOWLEDGEMENTS

The author thanks the Crossref team for providing free, open access to scholarly metadata via their REST API, and the open-source communities behind Streamlit and Plotly for their visualization frameworks.

## DECLARATION OF COMPETING INTEREST

The author declares that there are no known competing financial interests or any personal relationships that could have appeared to influence the work reported in this paper.

## DATA AVAILABILITY

The ScholarSearch system uses the Crossref REST API (https://api.crossref.org) as its sole data source. All data used in the evaluation is publicly accessible without authentication. The benchmark queries and evaluation scripts are available in the project's open-source repository under the MIT license.

# AUTHOR BIOGRAPHY

**Ivan Dobrovolskyi** is a Staff Software Engineer at Walmart Global Tech (Sunnyvale, CA), where he works on AI agent infrastructure for supply chain automation. Previously, he was a Staff Software Engineer at Blackhawk Network (Giftcards.com), where he led the development of ML-based content moderation systems. Earlier in his career, he served as Lead Architect for Public JSC 'Ukrainian Railways' (Ukrzaliznytsia) and led frontend development for eCherha, a digital queue platform for Ukrainian government services. He holds a Master of Science in AI Engineering from Western Governors University (2025–2026) and a Bachelor's degree in Computer Science from the National Technical University of Ukraine 'Kyiv Polytechnic Institute', SEC Institute for Applied System Analysis and Computer Systems Design (2012–2016). He is an IEEE Senior Member (SMIEEE). His research interests include agentic AI architectures, multimodal LLM systems, and computational approaches to content analysis.